\documentstyle[11pt]{article}
\begin{document}
%
 
\Large
{\bf 
Revealing Galaxy Associations in Abell 119
}
\vspace{0.2in}  
\large

V.G.~Gurzadyan$^*$ 
and A.~Mazure$^{\dagger}$
\vspace{0.1in}  
 
\vspace{0.2in}  
 
$^*$University of Sussex, Brighton, UK and
Department of Theoretical Physics, Yerevan Physics Institute,
Armenia (permanent address)\\
$^{\dagger}$IGRAP, Laboratoire d'Astronomie Spatiale, Marseille
\vspace{0.2in}  
 
We report the results of an analysis of the
hierarchical properties of the cluster of galaxies Abell 119.
Observational data from the ESO Nearby Abell 
Cluster Survey (ENACS) are used, complemented by data from previous studies, 
while the analysis is performed with the S-tree method.
The main physical system and its three subgroups with
truncated Gaussian velocity distributions  are identified;
due to their remarkable physical features and possible cosmogonical
mission we call these subgroups {\it galaxy associations}.
The mass centre of the core of main system is shown to coincide
with the X-ray centre of the cluster.
An alignment of the mass centres of the 3 galaxy associations is also shown.
\vspace{0.2in}  

The cluster of galaxies Abell 119 has attracted much attention, particularly
due to its remarkable X-ray properties, known since Ariel-V
measurements$^1$. Later observations by EXOSAT$^2$ allowed 
more detailed analysis of the physical conditions within the cluster, and,
combined with photometric and spectroscopic data for the
member galaxies, 
study of the substructuring properties of the cluster$^3$. The question
of substructure of cluster of galaxies is critical to an
understanding
of the mechanisms of their formation, and numerous studies
support the existence of subgroups in clusters
(see, {\em e.g.,}\/ ref.~4).
Below we reconsider the problem of the substructure of A119, 
based the ENACS data complemented with the data from ref.~3; this sample
includes information on the
redshifts, 2-D co-ordinates, and magnitudes of 142 galaxies. 
The ENACS dataset$^{5,6}$ contains redshifts 
for over 5000 galaxies
in 107 southern-sky clusters selected from the Abell-Corwin-Olowin 
catalogue.
The reliability and accuracy of the ENACS velocity measurements have been
extensively 
discussed in the presentation of the survey ({\em e.g.,}\/ Fig.~4  in ref.~6).
Specifically for
A119, comparison has been made between ENACS results and 
data by Fabricant {\em et al.}\/$^3$,
and no systematic errors have been found (see 
Table~2 in ref.~6).
\goodbreak
\medskip\noindent
{\it Method}.
We have performed an analysis of the hierarchical structure of A119
by means of the S-tree method; for details of this geometrical
technique we refer the reader to 
refs.~7--10. The method is based, essentially, on the
concepts of theory of dynamical systems, and particularly on the property of
structural stability, enabling the identification and study of robust
properties of nonlinear systems from limited amounts of
information. In the given problem, the method self-consistently takes
into account the positional, redshift, and magnitude information for the
galaxies.
The basic concepts of the method are the {\it degree of boundness}
between various members, and the corresponding
{\it S-tree diagrams,}\/ which represent the hierarchical
substructure of the system.
The problem is formulated such as to determine the correlation,
which should exist between the co-ordinates and velocities of
$N$ bodies if they are interacting gravitationally. 
It is performed by using the
degree of deviation of the trajectories of the system in $6N$-dimensional phase
space, by means calculating the two-dimensional curvature of that 
space$^{7,10}$:
$$
K^{\mu}_{\nu}=R^{\mu}_{\lambda\nu\rho}u^{\lambda}u^{\rho}.
$$
As described in the cited references, numerical
experiments show that the results of
the subgrouping typically are statistically significant (greater
than 90 per cent confidence)
when the total number of bodies is 
$N > 30$--35; the significance of the individually identified 
subgroups is the same,
since the dynamics of the whole system is being considered. The  method  has
been applied to the Local Group$^{10}$ and to 
the core the of Virgo cluster$^{11}$,
as well as for a sample of
ENACS clusters$^{12}$; their substructure was revealed, including the
membership of each individual galaxy in the subgroups and, in some cases,
morphological segregation between the subgroups have been noticed.
Let us briefly mention how the results of S-tree method
correspond to, say, those of wavelets$^{4}$. 
In collaboration with Eric Escalera, we have
performed a thorough comparative study of 
Abell-cluster data by both S-tree and wavelet techniques (see ref.~12).
The results are in fair agreement when defining  the
main system, but S-tree method is additionally able 
to reveal small-scale
subgroups, where the significance of the wavelets is limited.
This is not unexpected since, for wavelets, a giant galaxy and its satellite
have the same statistical weight while attracting the companions, while
S-tree also uses information on the magnitudes, by assuming
$M \propto L$; for details we again refer to refs.~7--9.
\goodbreak
\medskip\noindent
{\it Substructure}. 
The main results of the analysis are given in Table~1, which
includes parameters for the core of
main system (cMS) and the subgroups (1s, 2s, 3s).
The redshift histograms are shown in Fig.1. 
\begin{table*}
\centering
\caption{Derived parameters for Abell 119: the co-ordinates of the
mass centres of the core of main system (cMS) and of its subgroups; $N$ 
denotes the
number of galaxies; $m$ the median velocity;  and $\sigma$ the standard
deviation of redshift distribution.}
\medskip
\begin{tabular}{lllllll}
\hline
\hline
A119    & cMS         & 1s         &  2s        & 3s         & \\
\hline
centres & 00 53 31.84& 00 53 31.65& 00 53 34.03& 00 53 29.13& \\
        &-01 30 30.64&-01 31 33.29&-01 30 31.13&-01 24 27.80& \\
N       & 97         & 53         &  18        & 13         & \\
m       & 13173      & 13152      & 13702      & 12487      & \\
$\sigma$& 477        & 212        & 100        & 80         & \\
\hline
\end{tabular}
\end{table*}
The S-tree technique allows us to investigate further
the mutual degree of interaction of various subsystems or individual galaxies.
For example, our analysis does not attribute the group of galaxies at
$cz< 12 000$ to
the main system; its probable foreground nature has also been suggested
in ref.~3.
On the other hand, the analysis indicates the existence of a 
system of 12 galaxies, 
with mean redshift around 14$\,$200~km~s$^{-1}$, 
which probably has a physical connection with the
cMS, in that the S-tree diagram supports 
a weak correlation (with respect to the variation of
{\it degree of boundness} for cMS itself) between the cMS and
subgroup. 
The standard deviation, $\sigma$, for both systems (109 galaxies) presumably
situated in the same potential well, is $\sigma = 604$~km~s$^{-1}$.
If the cluster of galaxies is a more or less isolated system, then,
depending on the total number of galaxies and other initial conditions, its 
velocity distribution can be close to Gaussian, as we have found for the case 
of A119-cMS.
However, the subgroups -- {\it 'galaxy associations'} -- can have truncated
Gaussian distributions,
as it was shown in ref.~11, based on both observational data and
theoretical considerations. It was predicted$^{11}$ that
galaxy associations, while
moving through the host cluster, will inevitably lose their high-velocity
members, {\em i.e.,}\/ the `wings' of the Gaussian, so that the timescale for
the cut-off of the wings is smaller than that for their recovery.
It seems that A119 presents a similar case.
Quantitatively, this fact can be seen by estimating the 4th moment
of the velocity distribution (the kurtosis); this yields 
$-1.0$, $-1.0$, and $-1.1$ for
the 1s, 2s, and 3s subgroups, respectively, and 0.3 for the cMS.

\medskip\noindent
{\it Comparison with X-ray data.}
To probe the potential well,  X-ray data for the cluster 
are of particular importance. In the simplest case a single 
parameter, $\beta =\sigma^2\mu m_p/kT_x$, is often used to describe that 
correspondence. For A119 we have $\beta=0.68$, if the
{\it Einstein IPC}\/ value for the X-ray temperature is used$^{13}$, $T_{\rm X}
= 5.9$~keV
(see also ref.~14). As one can see from refs.~13 and 14, this
value of $\beta$ is not exceptional for clusters with known
X-ray temperatures, though one should recall that the content
of this parameter cannot be the same for a cluster with and without
substructure.
The mass centre of the cMS (Table 1) lies
within the centre of X-ray map of A119$^3$, thus supporting the hypothesis 
of a common potential well for the galaxies and the X-ray gas. 
X-ray observations with higher angular resolution can be of considerable
importance in resolving more-detailed structure of
the distribution of the X-ray gas, especially in view of the
existence of the subgroups and the possible hierarchical distribution of the
X-ray gas, and, therefore, of  dark matter$^{15}$. In that case, obviously,
the phenomenon should be described not by a single parameter $\beta$,
but by a sequence of more informative parameters corresponding to the cMS, 
to each of subgroups, and so on.

\medskip\noindent
{\it Centres of galaxy associations.}
Once the centres of mass of the subgroups -- galaxy associations -- are
obtained, their mutual location can be investigated;
this can conveniently be done with respect to the equation of a
line intersecting the three given points ($\alpha_i, \delta_i, i=1,2,3$):
$$
tg\delta_2\sin(\alpha_3-\alpha_2) + 
tg\delta_3\sin(\alpha_2-\alpha_1) +
tg\delta_1\sin(\alpha_1-\alpha_3)=0.  
$$
For the estimated co-ordinates of the 3 centres given in Table 1, the left-hand 
side of this equation yields $3.4{\times}10^{-6}$, thus implying 
good alignment of the projections of the centres. 

\medskip\noindent
{\it Conclusion}. 
We have performed an 
S-tree analysis of the substructure of A119.
The centre of core of the
main system of this cluster, extracted by the S-tree method from the total
sample of galaxies, coincides with the centre of its X-ray image,
{\it i.e.,}\/ with the region of maximum X-ray emission.
The substructure reveals 3 subgroups, the mass
centres of which are well aligned in projection.
In principle, chain-like structures might be expected from various 
formation mechanisms for
clusters of galaxies; however, more informative
descriptors should be involved 
for deeper study of
this phenomenon (see {\em e.g.,}\/ refs.~16, 17).
There is little value in comparing these 3 subgroups with the
subsystems mentioned in$^3$. Indeed, the
2-projection of galaxies up to 19$^m$ should, in the absence of information on 
their
redshifts, typically show `subsystems' having no actual connection to
A119, but being fractions of various projected clusters.
Thus no visible correlation correlation was noted in
ref.~3 between
the subsystems and X-ray image of A119, while our study has
revealed a correspondence with the X-ray data.
In general, the existence of box subgroups, i.e. galaxy associations, with
common parameters for
host clusters with a rather wide range of parameters$^{12}$, as we have
found for A119, could be interesting challenge for
various theories, since obviously they should have
primordial nature.
Observational study of the galaxies of the {\it galaxy associations}, such as
their morphology,
star-forming properties, certain features of the disk and the bulge of
spirals and so on, in comparison with other galaxies
of the cluster, can be of particular importance.  
We thank A.Melkonian for assistance with calculations.
V.G. was
supported by Royal Society and French-Armenian PICS.

\vfill\eject
{\it Figure caption}.
\centerline{\sc Fig. 1}
(a): The redshift histogram of the sample of 
Abell~119 galaxies
(solid line),
with the core of the main physical cluster, as extracted
by means of S-tree technique, indicated (dashed line).
\noindent
(b): The redshift histogram of the core of main cluster (solid line),
with the three identified box subgroups (dashed lines):
{\it 1s} is in the centre, {\it 2s} on the right, and {\it 3s} on the left.
\label{reconstruction}
\end{document}